\documentclass[fleqn,10pt]{wlscirep}

\title {Uncertainty principle for experimental measurements: Fast versus slow probes}

\author[1,2]{P. Hansmann}
\author[1,3]{T. Ayral}
\author[4]{A. Tejeda}
\author[1,5,6]{S. Biermann}

\affil[1]{Centre de Physique Th{\'e}orique, Ecole Polytechnique, CNRS, Univ. Paris-Saclay, 91128 Palaiseau, France}
\affil[2]{Max-Planck-Institut f\"ur Festk\"orperforschung, Heisenbergstrasse 1, 70569 Stuttgart, Germany}
\affil[3]{Institut de Physique Th{\'e}orique (IPhT), CEA, CNRS, URA 2306, 91191 Gif-sur-Yvette, France}
\affil[4]{Laboratoire de Physique des Solides, CNRS, Univ. Paris Sud, Univ. Paris-Saclay, 91405 Orsay, France}
\affil[5]{Coll\`ege de France, 11 place Marcelin Berthelot, 75005 Paris, France}
\affil[6]{European Theoretical Synchrotron Facility, Europe}

\begin{abstract}
The result of a physical measurement depends on the timescale of the
experimental probe. In solid-state systems, this simple quantum mechanical
principle has far-reaching consequences:  the interplay of several degrees of
freedom close to charge, spin or orbital instabilities combined with the
disparity of the time scales associated to their fluctuations can lead to
seemingly contradictory experimental findings. A particularly striking example
is provided by systems of adatoms adsorbed  on semiconductor surfaces  where
different experiments -- angle-resolved  photoemission, scanning tunneling
microscopy and core-level  spectroscopy -- suggest different ordering
phenomena. Using most recent first principles many-body techniques, we
resolve this puzzle by invoking the time scales of fluctuations when
approaching the different instabilities. These
findings suggest a re-interpretation of ordering phenomena and their
fluctuations in a wide class of solid-state systems ranging from organic
materials to high-temperature superconducting cuprates.
\end{abstract}

\begin{document}
\flushbottom
\maketitle

\thispagestyle{empty}

\begin{figure*}[t]   
  \begin{center}    
    \includegraphics[width=0.8\textwidth]{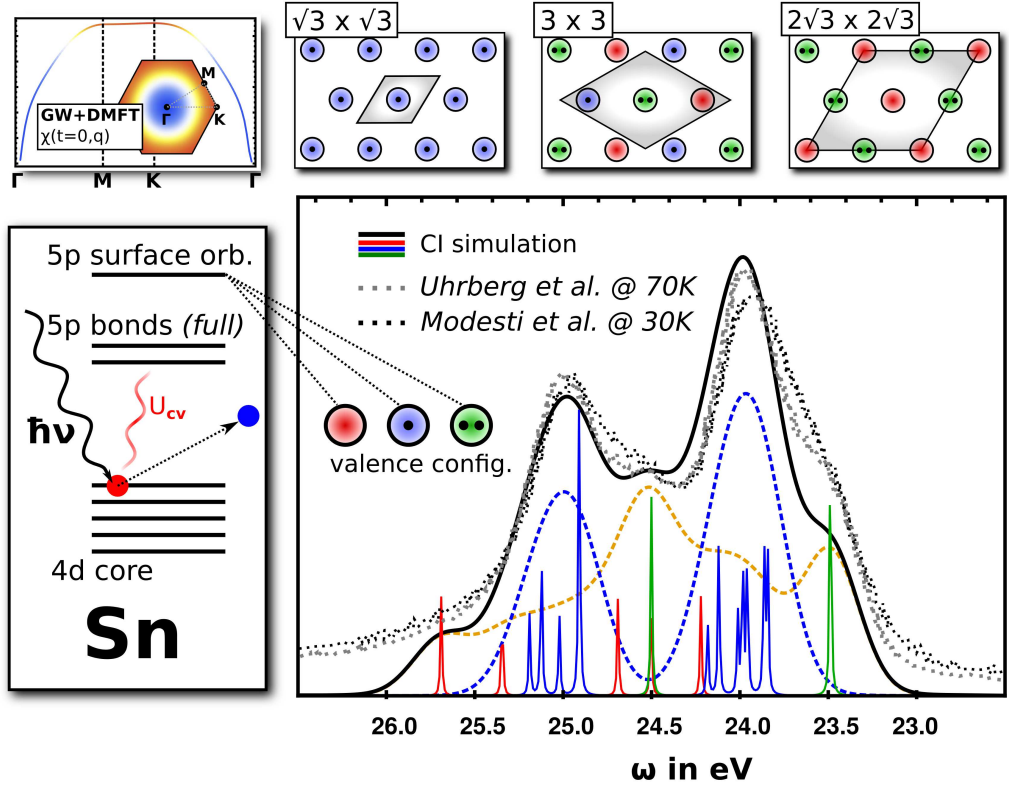}
    \caption{
      Core-level photoemission spectroscopy
      of the Sn adatom 2p-shell. Left hand side (top): GW+DMFT Charge
      susceptibility plotted allong the $\Gamma - M - K - \Gamma$ path in the
      Brillouin zone (see inset). Left hand side (bottom): Cartoon of the Sn
      4d core electron emission process. Right hand side (top): Sketches of
      the three surface configurations $(\sqrt{3}\times\sqrt{3})$R$30^{\circ}$,
      $3\times 3$, and  $(2\sqrt{3}\times2\sqrt{3})$R$30^{\circ}$. 
      Right hand side (bottom): comparison between
      experimentally obtained spectra (black and gray dots) and theoretical
      simulations with full multiplet cluster calculations (dashed and solid
      lines): The black solid line is the 
      final theoretical result broadened by a Gaussian of width $0.37$eV.
      It is the sum of the weighted contributions of the two coexisting phases
      close to the Mott-CO insulator transition (blue and orange dashed lines).  
      The solid narrow lines (narrow peaks) resolve the contributions to the
      total spectrum by empty surface orbitals (red), singly occupied surface
      orbitals (blue) and fully occupied surface orbitals (green)
      incorporating respective multiplet splittings.
    }
    \label{Fig1} 
  \end{center}
\end{figure*}
\begin{figure*}[t]   
  \begin{center}    
    \includegraphics[width=\textwidth]{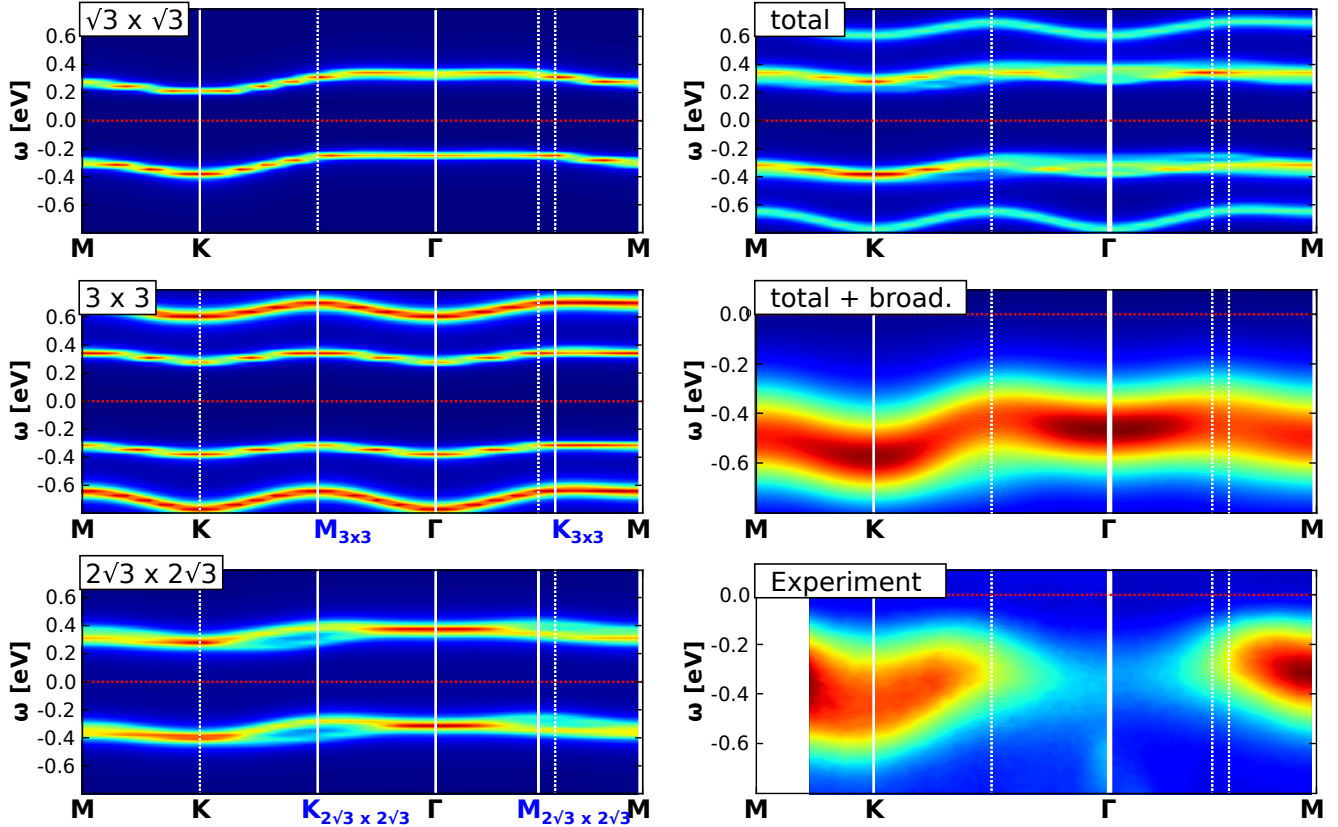}
    \caption{Left hand side panels: Correlated $A(\mathbf{k},\omega)$ simulations
      plotted along the $M \rightarrow K \rightarrow \Gamma \rightarrow M$
      path in the   $(\sqrt{3}\times\sqrt{3})$R$30^{\circ}$-Brillouin zone
      (high symmetry points of reconstructed phases are shown in blue) for the
      three relevant surface configurations   $(\sqrt{3}\times\sqrt{3})$R${30^{\circ}}$ (top), $3
      \times 3$ (middle), and   $(2\sqrt{3}\times 2\sqrt{3})$R$30^{\circ}$ (bottom) - note the
      backfoldings of the lower two spectral functions around the
      high-symmetry points of the corresponding Brillouin zones marked by
      white vertical lines and blue labels (For sketches of the respective unit cells see
      top panel of Fig.~\ref{Fig1}). The red dashed line marks the Fermi energy
      ($\varepsilon_F=0$).
      Right hand side panels: Weighted sum of $A(\mathbf{k},\omega)$ of the
      contributions shown on the left hand side (top). Electron removal part
      of the total spectral function with additional broadening (middle) for
      comparison with experimental ARPES data (bottom). Note that our
      simulation has no information about $\mathbf{k}$-dependent matrix
      elements of the actual ARPES measurement so that relative intensities of
      theory and experiment are not expected to be comparable.
      } 
    \label{Fig2} 
  \end{center}
\end{figure*}
\begin{figure*}[t]   
  \begin{center}    
    \includegraphics[width=\textwidth]{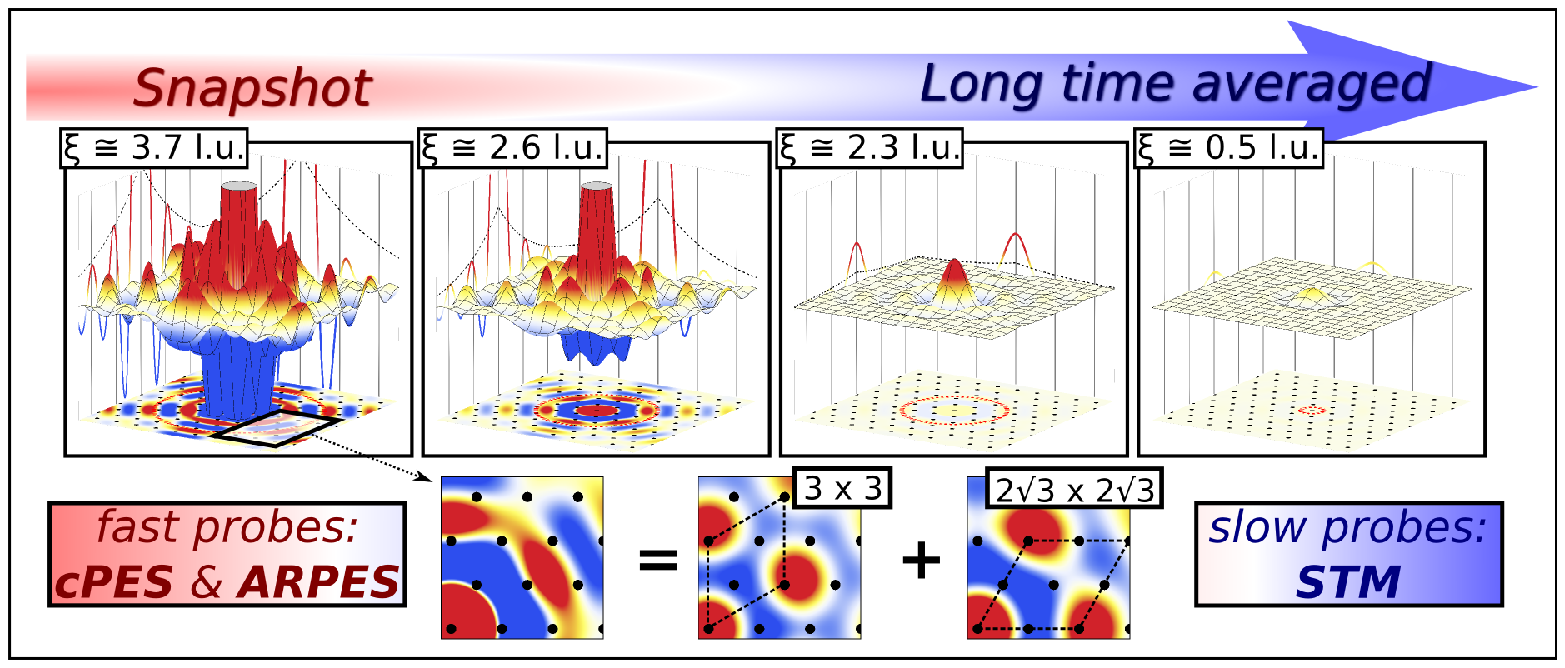}
    \caption{Upper panels: GW+DMFT charge susceptibility $\chi(\mathbf{R},\tau)$ plotted on
      the real space surface lattice (in the xy-plane - indicated by the black
      dots on the bottom of the respective plots) for four different values of $\tau$. At
      $\tau=0.0$ we find large charge fluctuations of correlation lengths
      $\xi$ exceeding $3.5$ lattice units (l.u.) which are picked up by
      core-level and photoemission sepctroscopies. Due to decay on a fs
      timescale (see evolution with $\tau$) they are invisible to slow probes
      like STM. Lower panels: The charge fluctuations can be decomposed into
      two dominant contributions related to $3 \times 3$ (``210'') and
        $(2\sqrt{3}\times 2\sqrt{3})$R$30^{\circ}$ (stripes) symmetry.}
    \label{Fig3}
  \end{center}
\end{figure*}

\section*{Introduction}

Understanding the mechanisms of charge, spin or orbital ordering and the
competition of different instabilities is a leitmotiv of modern solid-state
physics. Charge ordering phenomena in (quasi-)two-dimensional  transition
metal oxides have recently attracted tremendous attention \cite{Comin2014,
  Neto2014, Torchinsky2013, Comin2015, Tabis2014}.
Indeed, competing ordering phenomena might pave the way to superconductivity
and could be key to an understanding of the unusually high transition
temperatures observed in high T$_c$ cuprate superconductors. The competition
of different instabilities also leads to surprisingly complex phase diagrams
in a number of other materials, ranging from low-dimensional organics
\cite{BEDT1,BEDT2}, heavy fermion materials \cite{doniach77} to even simple oxides
\cite{pickett12, ralko15}, iridates \cite{takubo2014} or tellurides
\cite{hu2014}. 

Here, we study a material that epitomizes this interplay of
degrees of freedom, the timescales associated to their fluctuations and the
subtleties involved when probing them experimentally. $1/3$ of a monolayer of
Sn atoms adsorbed on the Si(111) surface -- a representative of the so-called
Si $\alpha$-phases -- forms a two-dimensional triangular lattice, where three
of the four valence electrons  of Sn are forming covalent bonds with the
silicon substrate while the remaining dangling bond results in a narrow
half-filled surface band. While the electronic structure at the band theory
level is thus extremely simple, many-body Coulomb correlations are a source
for complex phase diagrams which have been attracting considerable interest,
both experimentally
\cite{uhrberg85,grehk93,weitering93,carpinelli96,carpinelli97,
  weitering97,lay01,pignedoli04,upton05,modesti07,cardenas09,zhang10,
  tournier11,Cortes13,Li2013} and  theoretically
\cite{tosatti74,kaxiras90,brommer92,santoro99,hellberg99,
  aizawa99,profeta00,shi02,shi04,profeta05,profeta07,schuwalow10,chaput11,
  li11,Li2013}. Interestingly, despite the appealing single-orbital
nature of the physics of these compounds a simple model with  purely local
Hubbard interactions  falls short \cite{santoro99, hansmann_surfIOP,
  hansmann_surfPRL} of describing the observed phase diagrams. In particular,
charge ordering instabilities are driven by nonlocal interactions, and 
closely related compounds like Pb/Si(111) or Sn/Ge(111) are found to be in
symmetry-broken charge-ordered (CO) phases. For Sn/Si(111), experimental
results seem contradictory:  angle-resolved photoemission spectroscopy (ARPES)
shows backfoldings of bands associated to a 3x3 reconstruction of the unit
cell\cite{lobo03}, while scanning tunneling microscopy (STM) does not yield
any indication of any order. In a recent study Li \emph{et
al.} \cite{Li2013} proposed a magnetically-ordered state to be at the
origin of these contradictions. However, while the proposed spin order is
indeed a natural candidate, no direct experimental evidence for such an order
has been found yet. Moreover, as shown
below, only charge fluctuations solve an equally important puzzle raised by
core-level spectroscopies of Sn. There, local excitations of the Sn 4d core
shell suggest that the ground state of the Sn/Si(111) system is composed of
more than one Sn electronic configuration (i.e. valence states).  A similar
contradiction, also pointing towards the importance of \emph{charge}-degrees
of freedom, is found between ARPES and low-energy electron diffraction (LEED)
and has led to speculations about inhomogenous phases \cite{uhrberg2000}, with thermally
fluctuating Sn-positions \cite{carpinelli96, carpinelli97, Avila1999, Cortes13,
  Cortes06, Erwin2006}. 
On the other hand it has been shown\cite{hansmann_surfPRL} that Sn/Si(111) is in
immediate vicinity of a phase transition between a Mott insulating and a
charge-ordered phase. Here we show that this peculiar position is key to resolve the
above puzzles:
Close to the phase transition we find dynamic charge fluctuations (beyond
static approximations like LDA) of $3\times 3$ and  $(2\sqrt{3}\times
2\sqrt{3})$R$30^{\circ}$ symmetry with lifetimes on the order of femto
seconds.  This lifefime is long enough for probes like ARPES and core level
photoemission (cPES) to detect them while STM, as a static (i.e. time
averaging) probe, is blind to such dynamic modes. Our analysis becomes
quantitative when assuming a (likely) phase separation in vicinity of the
first order Mott-CO transition.      

\section*{Results}
Our starting point is the charge-charge correlation function computed within
combined many-body perturbation theory and dynamical mean field theory
(GW+DMFT), see upper left-hand side of  Fig.~\ref{Fig1}, reference
\cite{hansmann_surfPRL}, and the method section. 
This quantity suggests the presence of three major \emph{dynamic}
charge-fluctuation symmetries, which are represented in a cartoon-like fashion
in the upper right-hand side of Fig.~\ref{Fig1}):  a $3\times 3$
charge-ordered state, where three inequivalent sites are respectively doubly
occupied, half-filled and empty (a ``210'' charge distribution), a
$(2\sqrt{3}\times 2\sqrt{3})$R$30^{\circ}$ reconstructed state where stripes
of empty sites alternate with stripes of doubly occupied sites, and the
conventional Mott insulating state where  all sites are half-filled.  Its
time/frequency dependence indicates fluctuations of these symmetries with a
characteristic timescale (i.e. quasiparticles of a specific lifetime) of the
order of femto seconds, long enough for the cPES process to capture the
fluctuations in a ``snapshot''-like measurement. We remark that for  $3\times
3$ charge-ordered Sn/Ge(111) a charge distribution with one filled and two
quarter-filled sites per unit cell has been discussed. The configurations of
the many-body wave functions that would lead to such charge
distribution in the symmetry broken phase \emph{are} included in our GW+DMFT
calculations as charge fluctuations visible in our calculated charge
susceptibility. 

Mathematically, we determine the ground state wave function to be a
quantum superposition state (QS) composed of the three configurations with weights 0.13, 0.56 and 0.31 for
the $(\sqrt{3}\times \sqrt{3})$R$30^{\circ}$, $3\times 3$, and
$(2\sqrt{3}\times 2\sqrt{3})$R$30^{\circ}$ configurations respectively, see
Methods section.
With these insights, we now turn to a discussion of the
different observables measured in cPES, STM and ARPES.

\subsection*{Core-level spectroscopy} 
The first probe we consider is core electron emission from the Sn
4d-shell. Core-level spectroscopy is a local Sn probe as sketched on the
left-hand side of Fig.~\ref{Fig1} where a 4d core electron is emitted out of
the solid by an incoming photon. Due to the Coulomb interaction of this 4d
core hole with the 5p valence electrons (U$_\text{cv}$) the spectrum is
sensitive to the Sn valence configuration which consists of an either empty,
half-filled, or full surface orbital.  It is therefore a most efficient probe
for charge-ordered states or charge fluctuations on timescales slower than the
experimental process.  To arrive at a first principles description of the
core-level spectra we first determine the core-level emission spectra
corresponding to the three different valence configurations (blue, red and
green curves in the central panel of Fig.~\ref{Fig1} corresponding to singly
occupied, empty and full surface orbital configurations respectively) from
cluster multiplet simulations (see methods section). The main energy scale for
each contribution is the spin-orbit coupling of the core hole which splits
each spectrum into two main peaks associated to a core hole with total angular
momentum $J_\text{ch}=5/2$ or $J_\text{ch}=3/2$. On a smaller energy scale
(below experimental resolution), the core-valence interaction (U$_\text{cv}$)
leads to multiplet splittings within each $J_\text{ch}$ subspace (Note
that for a filled valence shell (green) and its fully spherical charge
distribution such multiplet splittings cannot occur). 
 
If the system were homogeneously in the QS state determined above
the resulting core-level spectra would be given by the superposition
of the spectra determined for the different valence states with contributions
of $\approx13\%$ of the   $(\sqrt{3}\times \sqrt{3})$R$30^{\circ}$ phase, $\approx56\%$ 
of the $3\times 3$ phase, and $\approx31\%$ of the   $(2\sqrt{3}\times
2\sqrt{3})$R$30^{\circ}$ phase.  Translated into Sn valence contributions this
corresponds to $\approx32\%$ half-filled and  $\approx68\%$
empty/doubly-occupied sites.  The resulting spectrum is plotted as
yellow-dashed line in  Fig.~\ref{Fig1}), and is found to give an
unsatisfactory description of the experimental data (black\cite{uhrberg2000}
/gray\cite{modesti07} dots).
If on the other hand, the system were in a pure Mott insulating
state, our estimate of the core-level spectrum would be given by
the contribution of the half-filled surface orbital only, 
broadened by the experimental resolution (blue dashed line in Fig.~\ref{Fig1}).
Obviously, this assumption does not hold either, confirming our analysis
of charge fluctuations contributing to the core-level spectrum.
Being an order to order transition, the Mott to CO transition is expected to
be of first order. This is confirmed by the behaviour of our charge
susceptibility (discussed further below). Phase separation is very likely in the vicinity of such a first-order
transition.

Indeed, closer inspection of the spectra corresponding to the two possible states and
comparison to the experimental spectra shows that while neither the
homogeneous QS state of the type determined above nor the Mott state yield
theoretical spectra in agreement with experiment, the sum of the two  spectra
with weights 0.7 for the QS state and 0.3 for the Mott state, does. Such an
incoherent superposition can be interpreted as simulating a spatial averaging
of the two phases, that is, a state where phase separation leads to a spatial
coexistence of Mott-insulating and charge-ordered patches with ratio 3/7. The
obviously good agreement with the experimental measurement gives support to
our interpretation of the sample being in an inhomogeneous state where Mott
insulating islands are  embedded into a dynamical QS background, and the
spatial averaging done by the core-level spectroscopy results in a weighted
average of the two contributions with weights 3:7. These results yield  a
quantum mechanical, complementary, perspective on speculations of an
inhomogeneous phase \cite{uhrberg2000}, albeit rather as a superposition of
the Mott state with the QS precursor we have described, locating Sn/Si(111) in
the phase coexistence region of a first-order phase transition between  these
two phases.\\ 

Comparison to the related compound   Sn/Ge(111) \textcolor{red}{\cite{Avila1999,
    Cortes06, Cortes13}} yields further insight:  experiments unambigously
find Sn/Ge(111) in a fully static charge-ordered phase of $3\times 3$
symmetry. In this case the phase coexistence has disapeared and the
QS patches have grown to macroscopic length scales at temperatures
below 60 K.  Moreover, for the static case the experimental timescale is
irrelevant and direct comparison of cPES and STM is unproblematic. In
Sn/Si(111), on the other hand, such comparison can only be made by considering
the snapshot-like nature of cPES measuring the spatial average of coexisting
phases. Before  turning to a quantitative discussion of the time and spatial
extent of the charge fluctuations, based on the charge-charge correlation
function $\chi_\text{GW+DMFT}$ as obtained from GW+DMFT (see methods section)
we revisit another experiment -- complementary to STM and cPES -- of
Sn/Si(111) that has caused recent controversies. 

\subsection*{Angle-resolved photoemission spectroscopy}
Just as in core-level spectroscopy, the comparison of ARPES with STM surface
images (suggesting absence of any charge order) is not straightforward. More
specifically, previous interpretations of ARPES spectra assumed the
breaking of some kind of spin- \cite{Li2013} or charge- \cite{lobo03}
symmetry of the ground state in order to account for backfolded features of
the momentum-resolved spectral function. The origin of such a symmetry breaking
has however been unclear so far, since there is no other experimental evidence
for it: a static charge-ordered state would contradict STM results, and no
experimental probe has found any direct indication for spin-ordering. In the
light of the previous discussion on the core-level spectroscopy and by
comparison of experimental ARPES spectra with theoretical calculations, we
will argue in the following that assuming such symmetry breaking is
unnecessary when taking into account the typical timescale of the experiment:
just like core-level spectroscopy, ARPES can be understood as a ``snapshot''
probe that spatially averages over the surface. Hence, short-lived charge
fluctuations of specific symmetry and finite but sufficient spatial 
extension (see next paragraph for details and quantification) will be picked up by ARPES and
incoherently averaged (such effect has been recently shown for AF spin
fluctuations and their impact on ARPES experiments for the high T$_c$ cuprates
\cite{wallauer15}). 

In order to provide a direct comparison between our theory and
experiment we have simulated the ARPES spectrum in the
$(\sqrt{3} \times \sqrt{3})$R$30^{\circ}$-, the  $3\times 3$- , and the
$(2\sqrt{3}\times 2\sqrt{3})$R$30^{\circ}$-phase. In Fig.~\ref{Fig2} we show
in the bottom right panel the experimentally obtained ARPES signal along a
specific path in the   $(\sqrt{3}\times \sqrt{3})$R$30^{\circ}$ Brillouin zone
(BZ) (taken from Ref.~\cite{Li2013}). 
Along with the experimental data, we also provide a complete map of
single-particle spectra, so as to illustrate how the total spectrum can be disentangled into its
components and provide reference spectra for future experiments on
Sn/Si(111) or related compounds in the X/Y(111) family (X being a group IV adatom and
Y a semiconductor like Si, Ge, SiC, etc.).

If we now take the mixture (and subsequent broadening) of the three
theoretically calculated spectra (left panels in in Fig.~\ref{Fig2}) with
weights determined above, the agreement between experimental spectra and theory becomes satisfactory. To
be precise, we can identify certain symmetry features, i.e. backfoldings, to
be related to a specific charge fluctuation.   The common feature of all shown
spectra is their insulating nature, i.e. a finite gap. In the   $(\sqrt{3}
\times \sqrt{3})$R$30^{\circ}$-phase (upper left) the gap separates an upper
and lower Hubbard band of the Mott insulating state In the $3 \times 3$ phase
(center left), we find (as expected from the $3 \times 3$ occupations) the
combination of a band-insulating (empty and doubly occupied sites) and
Mott-insulating (singly occupied sites) gap. Finally, the spectral function of
the $(2\sqrt{3}\times 2\sqrt{3})$R$30^{\circ}$-stripe phase (lower left)
separates the bands of the doubly occupied and empty lattice sites and, hence,
represents a band-insulating spectrum. While the momentum-structure of the
$(\sqrt{3} \times \sqrt{3})$R$30^{\circ}$ Hubbard bands closely resembles  the
dispersion of a free electron on the surface lattice, the CO phases display
characteristic backfolding features in the   $(\sqrt{3} \times
\sqrt{3})$R$30^{\circ}$ BZ. Particularly noteworthy are maxima of the spectral
functions along the $K-\Gamma$ and $M-\Gamma$ directions (attention must be
paid to the different conventions for the naming of BZ points in different
publications) that were subject to discussions in previous studies
\cite{lobo03, Li2013}. Finally, we remark that the difference of
the total energy scale (i.e. the Fermi level) between our simulation and
experiment ($\Delta \varepsilon_F \sim 0.15$-$0.2$eV) most probably originates from
i) uncertainties of the theory/experimental determination of the Fermi
level or ii) error bars of our ab initio calculated values for
onsite and non-local interactions.

\subsection*{Time and spatial resolution of the charge fluctuations} 
In the previous two paragraphs we have seen that spectroscopies such as core
electron emission and ARPES seem to suggest charge order and, hence, are in
contradiction to STM images of the   Sn/Si(111) surface. As alluded to before,
this contradiction can be resolved by considering the typical timescales of
the experiments: While the spectroscopies are spatially averaged but quasi
instantaneous snapshot probes, the STM complementarily yields time-averaged
but spatially resolved information. We will now report on the details of time
and spatial resolution of the charge fluctuations relevant to the phase
coexistence. Indeed, the  discussion on the different experiments above and
their interpretation is based on a result from our theoretical ab-initio
treatment of   Sn/Si(111) within self-consistent GW+DMFT applied to a low energy
Hamiltonian (see Methods section and Supplementary material). The 
most relevant quantity for the present discussion is the charge-charge correlation
function $\chi(\mathbf{q},\omega)$ resolved in momentum $\mathbf{q}$ and
frequency $\omega$ and its respective Fourier transforms. In our framework this
quantity is self-consistently obtained and can be employed as a sensitive
probe for charge-order instabilities. More specifically, the vicinity to a
transition into an ordered phase of a specific symmetry would be signaled by
the behaviour of  $\chi(\mathbf{q},\omega = 0)$ at the corresponding
$\mathbf{q}$ vector. Intuitive insight can be obtained from the Fourier
transformed correlation function in real space and time
$\chi(\mathbf{R},\tau)$ plotted in Fig.~\ref{Fig3}. With this quantity we can
find the typical correlation length $\xi$ and timescale $\Delta\tau_0$ of a
charge fluctuation. Since the Mott to CO phase transition is not a second
order transition, $\chi(\mathbf{R},\tau)$ does not become continuously long
range (i.e. $\chi(\mathbf{q},\omega = 0)$ does not diverge at a given
$\mathbf{q}$) but the correlation length $\xi\rightarrow\infty$ only increases
up to a finite value of the order of a few lattice constants before entering
the symmetry-broken phase.\\

In Fig.~\ref{Fig3} we plot the calculated
$\chi_\text{GW+DMFT}(\mathbf{R},\tau)$ for Sn/Si(111) on the z-axis at three
different imaginary time slices which correspond to averages over
increasingly large timescales. Four panels show the evolution from
instantaneous measurements to averages roughly over some femtoseconds. $x$- and
$y$- axes present the 2D surface indicated also by the black dots
at the respective adatom sites. From these plots we can conclude that the Mott
phase coexists with \emph{short-lived finite size lattice-commensurate charge
  fluctuations.} To quantify this claim we extract a correlation
length at $\tau=0$ of about  $~4\,l.u.$ (enough for backfolded bands to be
occupied in the ARPES experiment \cite{wallauer15}). These numbers immediately
resolve the spectroscopy vs. microscopy puzzle: Sn/Si(111) is found in close
vicinity but not yet in a charge-ordered phase (which can actually be reached
by substituting the Si substrate by a Ge one). However, the ordered phase is
preceded by quickly decaying charge fluctuations that can be picked up by
fast core-level and photoemission spectroscopies but not by STM.\\

\section*{Discussion}
In this work, we have demonstrated that the apparent contradictions between
STM, ARPES and core-level spectroscopy for two-dimensional systems of adatoms
adsorbed on semiconductor surfaces can be resolved by considering that i)
specific compounds like Sn/Si(111) are located in the phase coexistence
region of the first-order phase transition from a Mott insulator to a
charge-ordered insulator, and ii) the timescales intrinsic to the different
experiments matter: quickly decaying charge fluctuations (of specific
symmetries) can be seen by fast snapshot-like spectroscopies (core-level
spectroscopy, ARPES) while slow microscopy (STM) detects only a time-averaged
image in which the charge modulations are averaged out. We have shed light on
the history of controversial interpretations of Sn/Si(111) by quantifying
these statements, based on first principles many-body calculations using
combined many-body perturbation theory and extended dynamical mean-field
theory (GW+DMFT). In order to provide a direct theory-experiment comparison,
we have computed the observables of core-level spectroscopy and angular
resolved photoemission. Moreover, we have visualized and discussed the key
observable for dynamically fluctuating surface compounds: the charge
susceptibility. Our analysis underlines the need for a very careful analysis
of experimental results in circumstances where characteristic timescales of
the material (i.e. fluctuations) and the experimental probe
coincide.

Interestingly, in the related Sn/Ge(111) system, experimental
discrepancies at temperatures above the $3\times 3$ ordered phase have been
explained with single particle theories by a freezing-in of Sn-vibrations
along the surface normal \cite{Avila1999} while no such freezing-in is seen in
Sn/Si(111). It has been speculated that physics equivalent to Sn/Ge(111) might
occur only at lower temperatures \cite{uhrberg2000} which, however, has not been
confirmed so far. Other paradigmatic phase transitions have also been explained by dynamical
fluctuations, as in the metal-insulator transition on In/Si(111)\cite{Yeom99,
  Gonzalez06, Wippermann10} or in the  novel cluster-diffusion transition on
Sn/Si(111):B \cite{Srour15}. In all of these transitions, the fluctuations
have been argued to correspond to classical fluctuations  between inequivalent
configurations. While we do not exclude backcoupling to the lattice,
our study reveals that the experimental observations can be  explained from
the purely electronic \emph{many-body wavefunction}.

It is clear that this kind of phenomenology is not restricted to
adatom systems, but can be expected to occur quite generally in 
two-dimensional systems close to competing instabilities. 
Most notably, observations of charge-ordering fluctuations dominate 
the recent literature on high-temperature superconducting oxides
\cite{Tranquada1995, ghiringhelli2012, Neto2015}, 
with conflicting interpretations concerning stripe- or
checkerboard-type charge-ordering tendencies, their driving mechanism
and their implications for superconductivity.
These questions should be carefully reexamined in the light of our
findings. The present surface systems only provide
a particularly clean and tunable model system, without the complications
due to disorder, mixing in of multi-orbital or ligand degrees of freedom
present in the cuprates. The well-defined single-orbital character
of the surface systems allows for a truly first principles treatment,
providing us with a unique tool for refining our understanding
of competing instabilities in the proximity of different ordering
phenomena. Our results and conclusions do not negate
magnetic fluctuations (discussed by Li \emph{et al.} Ref.\cite{Li2013} and a very recent
study by Glass \emph{et al.} Ref.\cite{Li2013}) which should be considered as
complementary to the discussed charge fluctuations. Reconciling both, spin-
and charge degrees of freedom in one theoretical framework will be one of the
next challenges - fortunately the materials to be tested are real and
experimental data is accessible to support or falsify theoretical predictions.

\section*{Methods}
\subsection*{Charge correlation function}
The correlation function in the charge channel $\chi(\mathbf{R},\tau)$
displayed in Fig.~\ref{Fig3} has been obtained by spatial and temporal Fourier
transformation of the charge correlation function $\chi(\mathbf{k},
i\omega)$. The latter is computed from the polarization function
$P(\mathbf{k},i\omega)$ through the relation: 

\begin{equation} 
\chi(\mathbf{k},i\omega)  = \frac{-P(\mathbf{k}, i\omega)} {1 - v(\mathbf{k}) P(\mathbf{k}, i\omega)} 
\end{equation}

Here, $v(\mathbf{k})$ is the Fourier transform of the interactions
$v(\mathbf{R_i }- \mathbf{R_j}) = U \delta_{ij} + V \cdot
\frac{a}{|\mathbf{R_i} - \mathbf{R_j}|}$ ($a$ in the lattice constant and
$\mathbf{R_i}$ denotes a lattice site.). The polarization function is computed
in the GW+DMFT approximation \cite{GWDMFT, Biermann_2004, 
Sun2004, Ayral_PRB13},
namely as the sum of the impurity polarization $P_\mathrm{imp}(i\omega)$ and
of the nonlocal part of the bubble $\sim 2GG$, more specifically

\begin{equation} 
P(\mathbf{k}, i\omega)  = P_\mathrm{imp} (i\omega) + 2 \left[
  \sum_{\mathbf{q}, i\nu}   G(\mathbf{q}+\mathbf{k},i\omega+i\nu)
  G(\mathbf{q}, i\nu)   \right ]_\mathrm{nonloc}
\end{equation} 
where
$G(\mathbf{q}, i\nu)$ is the fully self-consistent Green's
function from a converged GW+DMFT calculation.

The factor of 2 stems from spin degeneracy. The values of the interaction
parameters are calculated within the constrained random phase approximation
\cite{hansmann_surfIOP,hansmann_surfPRL}, namely
$U=1.0$ eV and $V=0.5$ eV. A recent {\it ab initio} determination of the interaction parameters of the Si $\alpha$-phases (X/Si(111) with X=C, Sn, and
Pb) has found nonlocal interactions to be as large as 50$\%$ of the onsite
ones and established a materials trend with the Sn compound  being
``half-way'' between Mott insulating C/Si(111) and charge-ordered
Pb/Si(111)\cite{Brihuega2005}.

\subsection*{Ground-state wave function of charge-ordered state}
\begin{figure}[t]   
  \begin{center}    
    \includegraphics[width=0.5\textwidth]{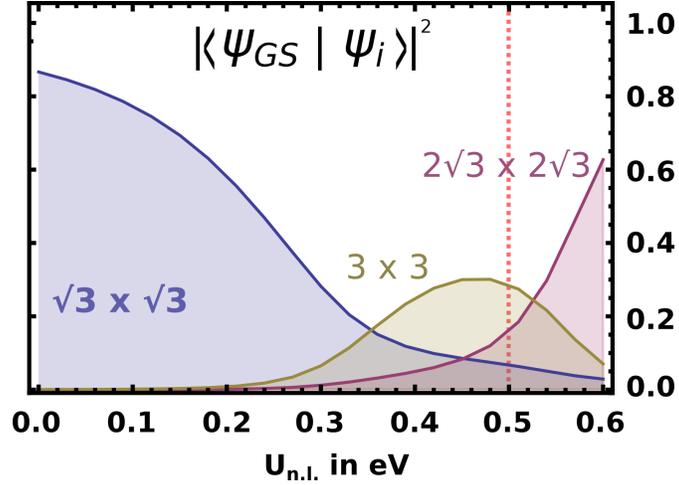}
    \caption{Projection of the many body ground state in a $6$-site cluster on
      its three most relevant contributions as function of the non-local
      interaction. The red dashed line indicates the cRPA value for Sn/Si(111).}
    \label{Fig4} 
  \end{center}
\end{figure}

In order to determine the weight with which the ``210'', stripe and Mott
configurations contribute to the ground-state wave function
in the charge-ordered state, we have solved --
by exact diagonalization -- a six-site cluster with periodic 
boundary conditions. Subsequent projection of the ground state on
the three relevant configurations of interest results in the 
estimates for the coefficients shown in Fig.~\ref{Fig4} as a function of non-local interaction.
The qualitative behaviour of the curves shown here can be understood as follows: For small
non-local interaction the Mott-like  $(\sqrt{3}\times \sqrt{3})$R$30^{\circ}$ phase is
dominant and unmixed with energetically high-lying configurations. However,
upon increasing 
the nonlocal interactions some of these high-lying configurations (in
particular the $3\times 3$ and   $(2\sqrt{3}\times 2\sqrt{3})$R$30^{\circ}$ states) become
lower in energy with respect to the ground state and for 
nonlocal interactions exceeding
$sim 0.34$ even replace the   $(\sqrt{3}\times \sqrt{3})$R$30^{\circ}$ configuration
as the main contribution for the ground state (note that mixing of the
different configurations is driven by the gain of kinetic energy,
i.e. electron hopping).

According to the cRPA calculations, the physical values for the nonlocal
interactions lead to a ground state composed of the above three 
components with coefficients 0.13 ($(\sqrt{3}\times \sqrt{3})$R$30^{\circ}$),
0.56 ($3\times 3$), and 0.31 ($(2\sqrt{3}\times 2\sqrt{3})$R$30^{\circ}$) -
normalized values extracted from 
the results shown in Fig.~\ref{Fig4}.
\\

\subsection*{Multiplet cluster calculations} In order to simulate the Sn
4d core-level spectra (shown in Fig.~\ref{Fig1}) we employ full multiplet
cluster calculations using the code introduced in Ref.~\cite{haverkort13}. As
common practice for such cluster simulated spectra we estimated the strength
of the core hole spin-orbit coupling (SOC), and the multipole part of the
core-valence interaction from atomic Hartree-Fock
calculations\cite{cowan70}. For valence and core SOC we use $\zeta_{5p}=0.40$
eV, and $\zeta_{4d}=0.41$ eV; for the multipole moments of the core valence
interaction we use the Slater integrals $F^1=0.46$ eV, $F^2=1.47$ eV, and
$F^3=0.42$ eV. The monopole part of the core-hole valence interaction is (in
combination with the onsite U of Sn) responsible for the relative shift of the
three spectra and fixed by the overall width of the spectrum. 
The spectral functions are calculated
with exact diagonalization in the cluster limit and broadened by convoluting
with a Gaussian of width $0.37$eV.\\

\subsection*{Theoretical ARPES spectra}
The theoretical ARPES spectrum (top-right and middle-left panels of
Fig.~\ref{Fig2}) is computed as a weighted average of the spectra
($A^\alpha(\mathbf{k},\omega) \equiv \frac{1}{N^\alpha_\mathrm{el}} \mathrm{Tr} \left[-\frac{1}{\pi}
  \mathrm{Im} \hat{G}^\alpha(\mathbf{k},\omega+i\eta) \right ]$) for the 3
symmetries $\alpha=\lbrace \sqrt{3}\times\sqrt{3}, 3\times3 , 2\sqrt{3}\times
2\sqrt{3} \rbrace$:

\begin{equation} 
A(\mathbf{k}, \omega) = {\sum_{\alpha} \lambda_\alpha
    A^\alpha(\mathbf{k},\omega)}/{\sum_{\alpha}
  \lambda_\alpha}
\end{equation} 

The relative weights $\lambda_\alpha$ are
determined from the cluster diagonalisation and
the cPES spectra (see above). $N^\alpha_\mathrm{el}$ denote the number of
electrons per unit cell for each symmetry (respectively 1, 3, 2).  In the
top-right panel, this spectrum is broadened with a Gaussian distribution of
mean deviation $\sigma=0.3$ to account for ARPES uncertainties. The individual
spectra are shown in the remaining three panels. The self-energy of the
  $(\sqrt{3}\times\sqrt{3})$R$30^{\circ}$ symmetry is obtained by MaxEnt analytical
continuation of the  imaginary-frequency impurity self-energy
$\Sigma_\mathrm{imp}(i\omega)$ computed self-consistently through an EDMFT
scheme~\cite{hansmann_surfPRL, Ayral_PRB13}.

We find this self-energy to be reminiscent of an atomic self-energy
with a renormalized interaction given by the self-consistently
computed effective impurity interaction $\mathcal{U}(\omega=0)$
as obtained from GW+DMFT.  For the $3\times 3$ symmetry, we
take the atomic self-energy for the half-filled band, while for the
empty and full bands we take the Hartree estimates for the self-energy: 
\begin{equation} 
\Sigma^{3\times3}(\omega) =
\left( \begin{array}{ccc}
\frac{U}{2}-3V & 0 & 0 \\
0 & \Sigma_{\rm imp.}(\omega)& 0 \\
0 & 0 & -\frac{U}{2}+3V \end{array} \right)
\end{equation}

For the $2\sqrt{3}\times 2\sqrt{3}$ symmetry, we also take Hartree estimates:
\begin{equation} 
\Sigma^{2\sqrt{3}\times 2\sqrt{3}}(\omega) =
\left( \begin{array}{cc}
\frac{U}{2}-2V & 0 \\
0  & -\frac{U}{2}+2V \end{array} \right)
\end{equation} 



\section{Acknowledgements}
We thank the authors of Ref.\cite{Li2013} and Daniel Malterre,
Yannick Fagot-Revurat, Alessandro Toschi, Georg Rohringer, Thomas Schaefer,
and Masatoshi Imada 
for useful discussions. We are particularly indebted to Ralph Claessen
and J{\"o}rg Sch{\"a}fer for providing us with the ARPES 
data of Ref.\cite{Li2013} and allowing us to replot them.
This work was supported by the French ANR under project
SURMOTT, IDRIS/GENCI under project 091393 and the European 
Research Council under its Consolidator Grant scheme 2013 
(project number 617196).

\section{Authors contributions statement}
PH performed the DFT, cRPA and cluster calculations. TA performed the GW+DMFT
calculations, TA and PH performed the calculations of the k-resolved spectra.
SB, AT and PH planned the study. All authors contributed to the
interpretation of the results and writing of the manuscript.

\section*{Additional information}
\textbf{Competing financial interests:} The authors declare no competing
financial interests.

\end{document}